\documentstyle[twoside]{article}

\newcommand{\cn}{\mbox{{\rm C}$\! \! \! $\raisebox{.45ex}{\tiny{$|$}}$\; $}}

\begin{document}
\begin{titlepage}

\vspace{0.5in}
\LARGE
\center{A stability analysis for the Korteweg-de Vries equation}
\Large
\vspace{0.5in}
\begin{center}
H.J.S. Dorren
\hspace{.5in}
and
\hspace{.5in}
R.K. Snieder
\end{center}
\vspace{5in}
\large
\center{\em Department of Theoretical Geophysics,
\\ Utrecht University,
\\ P.O. Box 80.021, 3508 TA Utrecht, The Netherlands}
\vspace{0.5in}
\end{titlepage}

\newpage
\begin{abstract}
In this paper the stability of the Korteweg-de Vries (KdV) equation is
investigated. It is shown analytically and numerically that small 
perturbations of solutions of the KdV-equation introduce effects of
dispersion, hence the perturbation propagates with a different velocity
then the unperturbed solution. This effect is investigated analytically
by formulating a differential equation for perturbations of solutions
of the KdV-equation. This differential equation is solved generally
using an Inverse Scattering Technique (IST) using the continuous part
of the spectrum of the Schr\"{o}dinger equation.
It is shown explicitly that the
perturbation consist of two parts. The first part represents the
time-evolution of the perturbation only. The second part represents the
interaction between the perturbation and the unperturbed solution. 
It is shown explicitly that singular non-dispersive solutions of the 
KdV-equation are unstable.
\end{abstract}

\section{Introduction}
Since the discovery of the IST,
a great number of nonlinear differential equations have been solved 
using
this technique
(see for instance ref.\cite{Ablowitz}
and the references therein). Most of the scientific effort has been
focused on finding soliton solutions for these equations. 
As firstly observed by Scott-Russell in 1834, 
solitons have the property that they maintain their shape over long
time-scales \cite{Scott}. 
From this observation the conclusion can be drawn that solitons which
are observed in nature are
stable solutions. 

In this paper we focus on the KdV-equation. The IST
approach, 
as introduced by Gardner, Greene, Kruskal and Miura (GGMT),
uses the
inverse problem of the Schr\"{o}dinger equation to solve the
KdV-equation. In their approach a
solution of the KdV-equation can be regarded as a time-dependent
potential of the Schr\"{o}dinger equation \cite{Gardner}.
GGMT have presented a method to compute the time-dependence of the 
S-matrix 
that is used to solve the inverse problem of the Schr\"{o}dinger
equation. Furthermore, they have shown that soliton solutions can be 
constructed using the discrete part of the spectrum of the 
Schr\"{o}dinger equation. 

Recently, it was discovered, (by  solving the KdV-equation 
using the continuous part of the spectrum of the Schr\"{o}dinger
equation) that  the KdV-equation has also singular solutions
\cite{Dorren0}.
It is also
shown in ref.\cite{Dorren0}, that the time-evolution of these singular
solutions can be associated
with a positive Lyapunov exponent. This implies that singular solutions
of the KdV-equation can exhibit unstable behavior.

It is remarkable that although the KdV-equation is one of the better
studied  equations in the field of mathematical physics, the matter of
the stability of this equation has not been investigated. The
classical argument that explains that solitons maintain their shape over
long time-scales is the presence of the nonlinear term in the
KdV-equation that cancels the effects of dispersion. However in
nature, solitons are always contaminated with noise and it is not clear
that when noise is added to a pure soliton solution, the soliton retains
its identity. This matter is related to the stability
properties of solutions of the KdV-equation. In this paper we
show that for the KdV-equation solitons posses stable behavior if the 
propagation velocity 
of the perturbation differs from the perturbation velocity of the
unperturbed soliton. 
In the case of the
singular solution this is also the case, but in contrast to the
soliton case, the amplitude of the perturbation can grow dramatically. 
This implies
that the shape of the singular solution changes. 

This paper has the following structure. In Sec.2 we discuss the balance
between nonlinearity and dispersion  for
non-dispersive
solutions of the KdV-equation. It is shown that for small
perturbations this balance is disturbed, and that the
perturbation has a different propagation velocity than the unperturbed
solution. In Sec.3, we derive a general expression for a perturbation
of a solution
of the KdV-equation.
It is shown that the time-evolution of the perturbation consists of two
parts. 
The first part represents the time-evolution of the perturbation in 
absence of the
unperturbed solution. The second part represents the interaction
between the soliton and the perturbation.
In Sec.4, we give an expansion of
the
time-evolution of the perturbation in a series solution. This enables
us to formulate an expression for the stability of solutions of the
KdV-equation. In Sec.5, an numerical example is given to illustrate
that the results that are derived in previous sections are also valid
in more general cases. The results are discussed in Sec.6. Technical
matter concerning the Gelfand-Levitan-Marchenko equations are added in
a appendix.

\section{Nonlinearity versus dispersion}
Consider the KdV-equation:
\begin{equation}
\begin{array}{l}
u_{xxx} - 6 u u_{x} + u_{t} = 0 \\
u(x,t=0) = u_{0}(x)
\label{kdv}
\end{array}
\end{equation}
In Eq.(\ref{kdv}), $u_{0}(x)$ represents the initial condition.
Non-dispersive solutions of Eq.(\ref{kdv}) can be obtained
by searching for solutions 
$u(x,t) = u(x-ct) \equiv u(z)$. 
In this special case Eq.(\ref{kdv}) reduces to:
\begin{equation}
u^{\prime \prime \prime} - 6 u u^{\prime} - c u^{\prime} =0
\label{kdv_nd}
\end{equation}
The KdV-equation 
(\ref{kdv}) is in the special case of non-dispersive solutions reduced
to a third-order ordinary nonlinear differential equation.  
In Eq.(\ref{kdv_nd}), the derivative $u^{\prime}$ stands for
$\frac{d}{dz} u(z)$. We can reformulate Eq.(\ref{kdv_nd}) in 
the following way:
\begin{equation}
\partial_{z} \left( 
u^{\prime \prime} - 3 u^{2}  - c u 
\right) = 0
\label{kdv_nd1}
\end{equation}
By integrating Eq.(\ref{kdv_nd1}), we find that
Eq.(\ref{kdv_nd}) is equivalent with:
\begin{equation}
u^{\prime \prime} - 3 u^{2} - cu +m =0
\label{kdv_in1}
\end{equation}
In Eq.(\ref{kdv_in1}), the constant $m \in I\!\!R$ is an
arbitrary integration constant. We can perform a further
simplification by multiplying both sides of Eq.(\ref{kdv_in1})
with a factor $u^{\prime}$. The total result can be formulated in 
the following way:
\begin{equation}
\partial_{z} \left( 
\frac{1}{2} ( u^{\prime} )^{2} - u^{3} - \frac{1}{2} c u^{2}
+ mu 
\right) = 0
\label{int1}
\end{equation}
By performing a further integration we find that Eq.(\ref{int1})
is equivalent with:
\begin{equation}
\frac{1}{2} ( u^{\prime} )^{2} - u^{3} - \frac{1}{2} c u^{2}
+ mu +n 
= 0
\label{kdv_in2}
\end{equation}
In Eq.(\ref{kdv_in2}), the constant $n \in I\!\!R$ is another
arbitrary integration constant.
In the special case that $m=n=0$, Eq.(\ref{kdv_in2}) is 
equivalent with:
\begin{equation}
u^{\prime} = \pm u \sqrt{2u+c}
\label{kdv_end}
\end{equation}
We have obtained that in the case of non-dispersive solutions the
KdV-equation (\ref{kdv}) is equivalent with Eq.(\ref{kdv_end}).
Eq.(\ref{kdv_end}) is an ordinary first-order nonlinear
differential equation that can be solved directly. This leads to
the following result:
\begin{equation}
u(z) =
\frac{
\frac{4 c}{D_{0}} 
e^{ \pm z \sqrt{c} }
}{
\left( 1 - \frac{2}{D_{0}} e^{ \pm z \sqrt{c} } \right)^{2}
}
\label{sol_rew}
\end{equation}
In  Eq.(\ref{sol_rew}) the constant $D_{0} \in I \!\!R$ can be
chosen arbitrarily and has to be determined by the initial condition.
If the constant $D_{0}$ is negative, the solution 
(\ref{sol_rew}) can be written as a soliton solution:
\begin{equation}
u(z) = - \frac{c}{2} \mbox{sech}^{2}(z \sqrt{c} + x_{0} )
\label{soluton}
\end{equation} 
If the constant $D_{0}$ is positive, this reduction does not
take place. It follows from Eq.(\ref{sol_rew}) that in this case
the solution of the KdV-equation has a singularity, but the solution 
propagates without dispersion. 
It has been argued in
ref.\cite{Dorren0} that these singular solutions can be associated with
a time-evolution having a positive Lyapunov exponent.
This leads to the conclusion 
that apart from the stable soliton solutions, the KdV-equation
has also unstable solutions.

Soliton solutions are understood to be stable solutions of the 
KdV-equation because the effects of dispersion and nonlinearity cancel 
each
other. It is tempting to choose Eq.(\ref{kdv_end}) as a 
starting point for the stability analysis for the KdV-equation since
Eq.(\ref{kdv_end}) is an ordinary differential equation on which
all the standard techniques of stability analysis can be applied.
However, due to the fact that Eq.(\ref{kdv_end}) is only valid
for non-dispersive solutions the corresponding stability analysis is
only valid for solutions of the KdV-equation (including a perturbation)
that maintain their shape. 
For non-dispersive solutions of the KdV-equation, the effects of
dispersion and non-linearity are in balance. If a perturbation is
added to a non-dispersive solution of the KdV-equation, the balance
between the dispersion and non-linearity is disturbed. 
In order to investigate this effect we
rewrite the KdV-equation in the following way:
\begin{equation}
u_{t} = - c u_{x} + ( c u_{x} - u_{xxx} ) + 6 u u_{x}
\label{disvnl}
\end{equation}
If the last two terms on the right-hand side of Eq.(\ref{disvnl})
are removed, we obtain:
\begin{equation}
u_{t} = -c u_{x}
\label{eqnl}
\end{equation}
Eq.(\ref{eqnl}) has the non-dispersive solution $u(x,t)=g(x-ct)$.
If the KdV-equation has non-dispersive solutions, the 
second and the third term on the right-hand side of 
Eq.(\ref{disvnl}) cancel each other. The $c u_{x} - u_{xxx}$ term on 
the right-hand
side of Eq.(\ref{disvnl}) describes the dispersion of the
solution $u(x,t)$. The $6 u u_{x}$ term on the right-hand side 
describes the
effects of the nonlinearity.

The solid line in Fig.1a the solid line represents a soliton at time 
$t=0$. The dashed line in Fig.1a, represents the soliton that is
contaminated with a $10 \%$ amplitude error. 
In Fig.1b, the balance between the effects of nonlinearity and the
effects of dispersion is shown  for the unperturbed soliton. 
The short-dashed line in Fig.1b represent the effect of dispersion
as given by the $c u_{x} - u_{xxx}$ term on the right-hand side of
Eq.(\ref{disvnl}). The long-dashed line describes the nonlinearity as
given by the $6 u u_{x}$  term on the right-hand side of
Eq.(\ref{disvnl}). The sum of these curves are given by the solid line
in Fig.1b. It follows from Fig.1b that the 
dispersion and nonlinearity are in balance. This  is consistent with
the fact that that the soliton propagates with velocity $c$ while
maintaining its shape.

If the soliton is contaminated with a small amplitude error, the
dispersion and nonlinearity are no longer in balance.
In Fig.1c, the examples of Fig.1b are repeated for the
contaminated soliton shown by the dashed line in Fig.1a. Similarly as in Fig.1b, the
short-dashed line in Fig.1c
represents the effect of dispersion as given by the $c u_{x} -
u_{xxx}$ term on the right-hand side of Eq.(\ref{disvnl}). The
long-dashed line describes the nonlinearity as given by the $6 u u_{x}$
term on the right-hand side of Eq.(\ref{disvnl}). The solid line in
Fig.1c represents the sum of the dispersion and the
the nonlinearity. An imbalance between
the dispersion and the nonlinearity is introduced due to the
amplification of the nonlinearity. As a result of the fact that the
nonlinearity and the dispersion no longer cancel, Eq.(\ref{eqnl})
(which is only valid for non-dispersive solutions of the KdV-equation),
does not describe the time-evolution of the perturbed solution.
The propagation effects of the
contaminated soliton is given by Eq.(\ref{disvnl}). The imbalance
between the effects of dispersion and nonlinearity is given by the solid
line Fig.1c, which is equal to the sum of the last two terms on the 
right-hand side
of Eq.(\ref{disvnl}).
In Fig.1d, the time-derivative of the perturbation is plotted. It turns
out that the time-derivative of the perturbation is not equal to a
constant times the space derivative of the perturbation. This implies
that the time-evolution of the perturbation is dispersive. In the
following sections this behavior is investigated analytically. 
Finally, in 
Fig.1e,
the finite difference solution of contaminated soliton is given
after at $t=2.2$ sec. It is observed that as a consequence of the
imbalance between the nonlinearity and dispersion, the perturbation
propagates with a different velocity than the unperturbed soliton,
and during the course of time, the contaminated soliton takes
its most natural form.

From this experiment, we can conclude that as a result of  the 
imbalance between the effects of nonlinearity and dispersion, the
perturbation of a solution of the KdV-equation propagates with a
different velocity. As a result of this the soliton and the perturbation
separate during the course of time. 

In Fig.2ab the examples of Fig.1 are repeated for the singular
solution. The solid line in Fig.2a describes the singular solution
of the KdV-equation at $t=0$. The short-dashed line in Fig.2a 
describes the
effects of dispersion as given by the $c u_{x} - u_{xxx}$
term on the right-hand
side of Eq.(\ref{disvnl}). The long-dashed line describes the
effects of nonlinearity as described by the $6 u u_{x}$  term on the
right-hand side of 
Eq.(\ref{disvnl}). Similarly as for the soliton, it turns out that also
for the singular solution effects of dispersion and nonlinearity are in
balance. If a small amplitude error is made, this balance is no longer
present. In Fig.2b, the effects of dispersion and nonlinearity are
plotted. The solid line in Fig.2b represents the sum of the 
nonlinearity and dispersion in the contaminated case.  It turns out
that the effects on nonlinearity and dispersion are no longer in
balance. As a result of this,  the perturbation propagates in the
opposite direction the unperturbed wave.

From the simple examples in this section, we can conclude that if
nondispersive solutions of the KdV-equation are contaminated with small
perturbations, dispersion effects are introduced.
The imbalance between the dispersion and nonlinearity is in the
coordinate frame that moves with the unperturbed solution very close to
the x-derivative  of the perturbation.
This implies that the perturbation has a non-zero velocity in this
reference frame so that the perturbation separates with time from the
unperturbed solution.

\section{ The stability of localized solutions}
The result in the previous section indicates that perturbations on the
initial condition of the KdV-equation propagate with a different 
velocity than the unperturbed solution.
In this section the effects of different perturbations are
investigated in a more general fashion.
If a perturbation $u(x,t) \rightarrow u(x,t) + f(x,t)$ is substituted
is Eq.(\ref{kdv}), the following
differential equation for the perturbation $f(x,t)$ can be derived:
\begin{equation}
\begin{array}{l}
f_{xxx}  - 6 \left( uf_{x} + f_{x}u + ff_{x} \right) + f_{t} = 0 \\
f(x,t=0) = f_{0}(x)
\end{array}
\label{d_pert}
\end{equation}
Eq.(\ref{d_pert}) represents a differential equation for the
perturbation $f(x,t)$, which depends on the unperturbed solution of the
KdV-equation $u(x,t)$. Eq.(\ref{d_pert}) can be solved using an
inverse scattering technique if a satisfying  Lax-pair is
constructed. However, because of the fact that the perturbed solution 
$u(x,t)+f(x,t)$ satisfies the KdV-equation,
the solution $f(x,t)$ of Eq.(\ref{d_pert}) 
can be computed directly
using the techniques described in ref.\cite{Dorren}. In Appendix A, a
brief overview of these methods is given:

As a starting point, we assume that the reflection coefficient
corresponding to the
unperturbed initial condition $u_{0}(x)$ undergoes a perturbation:
\begin{equation}
R(k,t=0) \rightarrow R(k,t=0) + \overline{R}(k,t=0) 
\label{ref_pert}
\end{equation}
In Eq.(\ref{ref_pert}), $R(k,t)$ describes the reflection
coefficient corresponding to the initial condition. Since the relation 
between the
reflection coefficient and the potential function is nonlinear, the
perturbation of the reflection coefficient $\overline{R}(k,t=0)$ 
cannot be associated with the spectral reflection coefficient 
corresponding to the
initial condition $f_{0}(x)$.
However, we can construct any initial condition $f_{0}(x)$, by 
imposing special conditions on the spectral reflection coefficient
$\overline{R}(k,t)$. With this we mean, the we can compute and analyze 
the time-evolution of different classes of perturbations $f(x,t)$, by
changing the analytical structure of $\overline{R}(k,t)$.
If both the unperturbed reflection coefficient and the perturbation of
the reflection coefficient are rational functions of the wave-number,
analytic expressions for $f(x,t)$ can be obtained.
Suppose $R(k,t)$ is a spectral reflection coefficient that can be
associated with the unperturbed solution $u(x,t)$. In Appendix A,
analytical expressions for the solution $u(x,t)$ are derived if the
reflection coefficient $R(k,t)$ is a rational function of the
wavenumber. From the unperturbed reflection coefficient $R(k,t)$, we
can construct a kernel $K(x,x,t)$ (Appendix A):
\begin{equation}
K(x,x,t) = \frac{
{\cal D}^{\prime}(x,t)
}{
{\cal D}(x,t) 
}
\label{kernel}
\end{equation}
In Eq.(\ref{kernel}) stands the prime for the derivative with
respect to the space-coordinate $x$.
The determinant ${\cal D}(x,t)$ in Eq.(\ref{kernel}) is given by:
\begin{equation}
{\cal D}(x,t) = \mbox{det} \left\{ \delta_{ij} - (p_{i}+p_{j})^{-1}
R_{j} e^{2i(p_{j}x + 4p_{j}^{3}t)} \right\}
\label{deter}
\end{equation}
In Eq.(\ref{deter}), $p_{i}$ are the poles and $R_{i}$ the
residues of the unperturbed reflection coefficient $R(k,t)$
Solutions of the KdV-equation can be derived by taking the following
derivative:
\begin{equation}
u(x,t) = -2 \frac{d}{dx} K(x,x,t)
\label{sol_kdv}
\end{equation}
If the reflection coefficient $R(k,t)$ is contaminated with a small
perturbation $\overline{R}(k,t)$, Eq.(\ref{deter}) contains
the poles
and residues of both the unperturbed reflection coefficient
$R(k,t)$ and the reflection coefficient corresponding to the 
perturbation 
$\overline{R}(k,t)$.
It is shown in ref.\cite{Dorren}, that if the reflection coefficient 
$R(k,t)$ undergoes a perturbation as given in Eq.(\ref{ref_pert}),
the determinant ${\cal D}(x,t)$ can be expanded into the following
series:
\begin{equation}
{\cal D}(x,t) \rightarrow {\cal D}(x,t) + {\cal E}(x,t)
\label{detpert}
\end{equation}
In Eq.(\ref{detpert}), ${\cal D}(x,t)$ is the determinant
(\ref{deter}) in absence of perturbations. The effect of the
perturbation is expressed in the determinant ${\cal E}(x,t)$. 
If Eq.(\ref{detpert}) is substituted in 
Eq.(\ref{kernel}) we obtain the following result:
\begin{equation}
K(x,x,t) \rightarrow 
\frac{ 
{\cal D}^{\prime}(x,t) + {\cal E}^{\prime}(x,t) 
}{
{\cal D}(x,t) + {\cal E}(x,t) 
}
\label{kerpert}
\end{equation}
Using some basic algebra, the kernel associated with the
unperturbed solution $u(x,t)$ can be separated from the right-hand 
side of Eq.(\ref{kerpert}): 
\begin{equation}
K(x,x,t)  \rightarrow 
\frac{ {\cal D}^{\prime}(x,t)
}{
{\cal D}(x,t)
} + 
\frac{
{\cal D}(x,t) {\cal E}^{\prime}(x,t) -
{\cal D}^{\prime}(x,t) {\cal E}(x,t)
}{
{\cal D}(x,t) [ {\cal D}(x,t) + {\cal E}(x,t) ]
} 
\label{xxxx}
\end{equation}
The term ${\cal D}^{\prime}(x,t)/{\cal D}(x,t)$ in Eq.(\ref{xxxx}) 
can be identified with the time-evolution of the unperturbed problem.
From Eq.(\ref{xxxx}), we can identify an expression
for the perturbation $f(x,t)$:
\begin{equation}
f(x,t)  =  -
2 \frac{d}{dx}  \left\{
\frac{
{\cal D}(x,t) {\cal E}^{\prime}(x,t) -
{\cal D}^{\prime}(x,t) {\cal E}(x,t)
}{
{\cal D}(x,t) [ {\cal D}(x,t) + {\cal E}(x,t) ]
} 
\right\}
\label{solupert}
\end{equation}
It should be realized that in the determinant ${\cal E}(x,t)$ 
the poles and residues of both the reflection coefficients
$R(k,t)$ and $\overline{R}(k,t)$ are
present. It follows from Eq.(\ref{xxxx}) that the perturbation
$f(x,t)$
is large with respect to the unperturbed solution if the denominator ${\cal D}(x,t)[
{\cal D}(x,t) + {\cal E}(x,t)]$ of Eq.(\ref{solupert}) is small.
As illustrated in the following examples, in this
case the perturbation $f(x,t)$ can dominate the total solution of the
KdV-equation.

In the following examples, we consider the case in which the unperturbed
solution of the KdV-equation has one single pole ($p=i \beta$) and one 
residue ($R=id$). The unperturbed solution of the KdV-equation has
either soliton-like behavior as in Eq.(\ref{soluton}) or singular 
behavior depending on the
position of the pole and the residue. In the following experiment, we 
choose
in case of the soliton $d=-1$ and $\beta =1$. We can illustrate the
effects of perturbations on the soliton by adding a certain number of
poles and residues to the unperturbed determinant ${\cal D}(x,t)$.
In the lower panel of Fig.3a,
the time-evolution of a contaminated soliton $u(x,t)$ is plotted.
As a reference, in the upper panel of Fig.3a the time-evolution of the
unperturbed soliton is given. It
can be observed from Fig.3a that the effects of the  contamination
either spreads out, or travel with a different velocity. This
implies that after a certain amount of time the unperturbed
soliton and the effects of the perturbation separate. This is more clear
in Fig.3b. The short-dashed line in Fig.3b indicates the initial
condition and the solid line indicates  unperturbed initial condition.
The long-dashed line indicates the solution at $t=1$. 
In the example of Fig.3, the positions of the poles and residues is
chosen in such a way that the numerical value of the denominator of
Eq.(\ref{solupert}) does not differ significantly from the
numerical value of ${\cal D}(x,t)$.
As a result of
this the perturbation remains in the same order of magnitude as the
unperturbed solution. 

The situation dramatically changes if we chose 
$d=0.01$ and $\beta =1$ and keep the positions of all the other 9 poles
and residues which specify the perturbation constant. By choosing 
$d=0.01$ and $\beta =1$, the
unperturbed  determinant ${\cal D}(x,t)$ can by equal to zero for
certain values of $x$ and $t$. Because of the analytical structure of
the determinant (\ref{detpert}), the perturbation ${\cal E}(x,t)$ has
in this case also zeros for certain values of $x$ and $t$. 
The time-evolution is in this case plotted in the lower panel of 
Fig.4. This result has to be compared with the time-evolution of the
unperturbed case which is given in the upper panel in Fig.4.
It follows that in this special case, the perturbed solution consists
of two different branches. The first branch
propagates with a similar propagation velocity as the unperturbed
solution but it has large amplitude fluctuations on the
characteristic of the unperturbed solution. The second part propagates
with a different velocity than the unperturbed solution. In the
following section, we derive analytical expressions for these different
parts of the perturbation. We want to remark that a small perturbation
of a singular solution at $t=0$ can always be constructed by choosing the
positions of the poles and residues properly. It follows from the
structure of equation (\ref{solupert}) that these small perturbations
can grow when ${\cal D}(x,t)[{\cal D}(x,t) + {\cal E}(x,t)]$ is
small.

From the  results in this section we can conclude that the soliton
exhibits stable behavior whereas the singular solution is unstable. 
Eq.(\ref{solupert}) gives a general expression for the 
perturbation of a solution of the KdV-equation. The perturbation
is small with respect to the unperturbed solution, if the unperturbed
solution has no poles close to the origin in the complex plane since
the dominator ${\cal D}(x,t) + {\cal E}(x,t)$ can not be zero. In
contrast to this, the perturbation is large with respect to the 
unperturbed solution if the dominator ${\cal D}(x,t) + {\cal E}(x,t)$ 
in Eq.(\ref{solupert}) is nearly singular. This is the case if the
unperturbed problem has poles close to the origin in the complex
plane.
This means that the singular solution is sensitive for noise. 
Moreover, for both the soliton and the singular solution the
propagation velocity of the perturbation, which determines whether the
perturbation separates from the unperturbed solution, is a crucial 
parameter for the stability. The propagation velocity and the amplitude
of the perturbation depend of the position of the poles and residues of
the  perturbation. 
In the
following section we study this in
more detail.

\section{A series solution for $f(x,t)$}

In order to analyze the behavior of the perturbation $f(x,t)$ it is
convenient to formulate this function as a series solution. 
Our starting point is the Marchenko equation without bound states 
in the wave-number domain as given in Appendix A. 
If the reflection coefficient undergoes a
perturbation (\ref{ref_pert}), we find the following relation.
\begin{displaymath}
F(k,x,t)= 
1+ \frac{1}{2 \pi i} \lim_{\epsilon \rightarrow 0+}
\int_{-\infty}^{\infty} dk^{\prime}
\frac{ 
\left\{ R(k^{\prime},t=0) + \overline{R}(k^{\prime},t=0) \right\}
F(k^{\prime},x,t)exp[2i(k^{\prime}x + 4 \{ k^{\prime} \}^{3} t)]
}{
k^{\prime} + k + i \epsilon
}=
\end{displaymath}
\begin{equation}
1+ \int_{-\infty}^{\infty} C(k,k^{\prime},t) F(k^{\prime},x,t)
dk^{\prime}+
\int_{-\infty}^{\infty} \overline{C}(k,k^{\prime},t) F(k^{\prime},x,t)
dk^{\prime}
\label{fexp_c}
\end{equation}
The function $F(k,x,t)$ is related to the kernel
$K(x,y,t)$ by the following Fourier transform: 
\begin{equation}
K(x,y,t)=(2\pi)^{-1} \int_{-\infty}^{\infty}
dk e^{-ik(y-x)}( F(k,x,t) -1 )
\label{kertran_c}
\end{equation}
Furthermore, 
the kernel $C(k^{\prime},k,t)$ in Eq.(\ref{fexp_c}) is given 
by:
\begin{equation}
C(k,k^{\prime},t) = \lim_{\epsilon \rightarrow 0+}
\frac{1}{2i\pi} \frac{
R(k^{\prime},t=0) e^{2i(k^{\prime}x + 4 \{ k^{\prime} \}^{3} t )}
}{
k^{\prime} + k + i \epsilon
}
\label{cker_c}
\end{equation}
The kernel $\overline{C}(k^{\prime},k,t)$ in Eq.(\ref{fexp_c}) is defined
by:
\begin{equation}
\overline{C}(k,k^{\prime},t) = \lim_{\epsilon \rightarrow 0+}
\frac{1}{2i\pi} \frac{
\overline{R}(k^{\prime},t=0) e^{2i(k^{\prime}x + 4 \{ k^{\prime} \}^{3} t )}
}{
k^{\prime} + k + i \epsilon
}
\label{cker1_c}
\end{equation}
Eq.(\ref{fexp_c}) can be represented schematically using a Dyson's
representation as given if Fig.5. Using an iteration technique, the
Dyson's series of Fig.5, can be expanded. The result is
given in Fig.6. It is observed from Fig.6, that the total
expression for the function $F(k,x,t)$ consists of three parts. 
The first part can be identified with all diagrams consisting of solid 
dots only. This series of diagrams represents the time-evolution of
the unperturbed solution. 
The second series consists of diagrams having open dots only. This
series of diagrams represents the time-evolution the perturbation in
absence on the unperturbed solution.
The remaining diagrams consist of a combination of both solid and open 
dots. This
series of diagrams represents the interaction between the unperturbed
solution and the perturbation. We can formally solve the Dyson's 
equation by iteration. The solution is shown by the diagrams in 
Fig.6, and is given by:
\begin{displaymath}
F(k,x,t) = 1 + \int_{-\infty}^{\infty} 
C(k^{\prime},k,t) dk^{\prime}
+ \int_{-\infty}^{\infty} \int_{-\infty}^{\infty}
C(k,k^{\prime},t)C(k^{\prime},k^{\prime\prime},t ) dk^{\prime} 
dk^{\prime \prime} + \cdots
\end{displaymath}
\begin{displaymath}
+ \int_{-\infty}^{\infty} \overline{C}(k^{\prime},k,t) dk^{\prime}
+ \int_{-\infty}^{\infty} \int_{-\infty}^{\infty}
\overline{C}(k,k^{\prime},t)
\overline{C}(k^{\prime},k^{\prime\prime},t) dk^{\prime} 
dk^{\prime \prime} + \cdots
\end{displaymath}
\begin{equation}
+ \int_{-\infty}^{\infty} \int_{-\infty}^{\infty}
C(k,k^{\prime},t)\overline{C}(k^{\prime},k^{\prime\prime},t) 
dk^{\prime} 
dk^{\prime \prime} + 
\int_{-\infty}^{\infty} \int_{-\infty}^{\infty}
\overline{C}(k,k^{\prime},t)C(k^{\prime},k^{\prime\prime},t) dk^{\prime} 
dk^{\prime \prime} + \cdots
\label{tot_ser}
\end{equation}
If both $R(k,t=0)$ and $\overline{R}(k,t=0)$ are rational functions of the
wave-number, the integrations in the series (\ref{tot_ser}) can be
carried out analytically 
by performing a contour integration in
$\cn^{+}$. This is justified by the fact that the reflection
coefficient $R(k,t=0) \rightarrow {\cal O}(1/k)$ if
$k \rightarrow \infty$.
The poles of the denominator of Eq.(\ref{tot_ser}) are all 
situated
in $\cn^{-}$ so the only contribution to the integrals of 
Eq.(\ref{tot_ser}) comes from the poles of $R(k,t=0)$
which are situated in $\cn^{+}$.
We write Eq.(\ref{tot_ser}) in the following way:
\begin{equation}
F(x,k,t) - 1 = F_{un}(x,k,t) + F_{pe}(x,k,t) + F_{int}(x,k,t)
\label{all}
\end{equation}
The function $F_{un}(x,k,t)$ only contains contributions of $R(k,t)$
and
can be identified with the time-evolution of the unperturbed solution
(solid-dot diagrams).
The function $f(x,t)$ consists of contributions of both $F_{pe}(x,k,t)$
and $F_{int}(x,k,t)$. 
The term $F_{pe}(x,k)$ only contains contributions of $\overline{R}(k,t)$ and
can be identified with the time-evolution of the perturbation in
the absence of $u(x,t)$ (open-dot diagrams). 
The term $F_{int}(x,k,t)$ contains contributions of both $R(k,t)$ and
$\overline{R}(k,t)$ and represents the interaction between the unperturbed
solution $u(x,t)$ and the perturbation (diagrams consisting of
solid-dots and open-dots). In the following we analyze 
these three terms separately.

\subsection{The time-evolution of the non-interaction terms}
Suppose the unperturbed reflection coefficient $R(k,t)$ has $N$
poles. It follows from Eq.(\ref{tot_ser}) and Eq.(\ref{all}) that the
contribution 
to the function $F(k,x,t)$ from the terms containing only the reflection
coefficient $R(k,t)$ (solid-dot diagrams) is equal to:
\begin{displaymath}
F_{un}(x,k,t) = \sum_{i=1}^{N} \frac{R_{i}}{k+p_{i}} 
e^{2i(p_{i}x +4p_{i}^{3}t)} +
\sum_{i,j=1}^{N} \frac{R_{i}R_{j}}
{(k+p_{j})(p_{i}+p_{j})}
e^{2i \{ (p_{i}+p_{j})x + 4(p_{i}^{3} + p_{j}^{3})t \} }
\end{displaymath}
\begin{equation}
\sum_{i,j,l=1}^{N} \frac{R_{i}R_{j}R_{l}} 
{(k+p_{l})(p_{j}+p_{l})(p_{i}+p_{j})}
e^{2i \{ (p_{i}+p_{j}+p_{l})x + 4( p_{i}^{3} + p_{j}^{3} +p_{l}^{3})t
\} }
\cdots,
\end{equation} 
where $p_{i}$ and $R_{i}$ are the poles and residues of the
unperturbed reflection coefficient.
If the Fourier transform (\ref{kertran_c}) is now performed on the
unperturbed part of $F_{un}(x,k,t)$, we obtain the following expression 
for the kernel $K_{un}(x,y,t)$:
\begin{displaymath}
K_{un}(x,y,t) = i \sum_{i=1}^{N} 
R_{i}
e^{2ip_{i}(x+y) +8ip_{i}^{3}t} +
i \sum_{i,j=1}^{N} \frac{R_{i}R_{j}}{p_{i}+p_{j}}e^{ip_{j}(x+y)}
e^{2ip_{i}x}
e^{8i(p_{i}^{3} + p_{j}^{3})t}
+
\end{displaymath}
\begin{equation}
i \sum_{i,j,l=1}^{N} 
\frac{R_{i}R_{j}R_{l}}{(p_{j}+p_{l})(p_{i}+p_{j})}
e^{ip_{l}(x+y)}e^{2i(p_{i}+p_{j})x}
e^{8i(p_{i}^{3} + p_{j}^{3} + p_{l}^{3} )t}
+ \cdots
\label{ker_ser_c}
\end{equation}
After putting $y=x$ and taking the derivative:
\begin{equation}
u_{un}(x,t)  = - 2  \frac{d}{dx} K_{un}(x,x,t), 
\end{equation}
the following expression for the unperturbed solution is obtained.
\begin{displaymath}
u_{un}(x,t)= 4 \sum_{i=1}^{N}  R_{i}p_{i} e^{2i(p_{i}x+4p_{i}^{3}t)} +
4 \sum_{i,j=1}^{N} R_{i}R_{j}e^{2i \{ (p_{i}+p_{j})x + 4(p_{i}^{3} +
p_{j}^{3} ) t \}  } +
\end{displaymath}
\begin{equation}
4 \sum_{i,j,l=1}^{N} 
\frac{(R_{i}R_{j}R_{l})(p_{i}+p_{j}+p_{l})}{(p_{i}+p_{j})(p_{j}+p_{l})}
e^{2i \{ (p_{i}+p_{j}+p_{l})x + 4(p_{i}^{3} +p_{j}^{3} + p_{l}^{3})t
\}}
+ \cdots
\label{part_un}
\end{equation}
This result is already obtained in ref.\cite{Dorren0}. From this
result we can conclude that a general solution of the KdV-equation can
be expanded in an infinite series of exponential basis functions. 

In a similar manner we can derive the time-evolution  of the
contributions to the Dyson's series in Fig.6 for all the terms that
can
be identified with the perturbation only (open-dot diagrams). 
Suppose the perturbation on the reflection coefficient 
$\overline{R}(k,t)$ 
has $M$ poles, it follows using a similar argumentation as for the
evaluation of the term $u_{un}(x,k,t)$ that:
\begin{displaymath}
u_{pe}(x,t)= 4 \sum_{i=1}^{M} \overline{R}_{i}
\overline{p}_{i} e^{2i(\overline{p}_{i}x+4\overline{p}_{i}^{3}t)} +
4 \sum_{i,j=1}^{M}
\overline{R}_{i}\overline{R}_{j}
e^{2i \{ (\overline{p}_{i}+\overline{p}_{j})x + 
4(\overline{p}_{i}^{3} +
\overline{p}_{j}^{3} ) t \}  } +
\end{displaymath}
\begin{equation}
4 \sum_{i,j,l=1}^{M} 
\frac{
( \overline{R}_{i} \overline{R}_{j} \overline{R}_{l} )(
\overline{p}_{i}+ \overline{p}_{j}+ \overline{p}_{l})
}{
(\overline{p}_{i}+\overline{p}_{j})
(\overline{p}_{j}+\overline{p}_{l})}
e^{2i 
\{ (\overline{p}_{i}+ \overline{p}_{j}+ \overline{p}_{l})x + 
4(\overline{p}_{i}^{3} + \overline{p}_{j}^{3} + \overline{p}_{l}^{3})t
\}}
+ \cdots
\label{part_pert}
\end{equation}
We observe
from this result that both the unperturbed solution and the
time-evolution of the non-interaction elements have a similar
analytical structure. We remark that the solution $u_{un}(x,t)$ 
corresponds to a series expansion of the terms in 
Eq.(\ref{xxxx}) is which only the determinant ${\cal D}(x,t)$ is
present. 
We can conclude that both the unperturbed
solution and the perturbation evolve as an infinite series of
non-dispersive solutions in time. However, the
spectral components of the perturbation $u_{pe}(x,t)$ generally travel
with a different
velocity than $u_{un}(x,t)$. 
This is already observed  in Fig.1 and Fig.2. In Fig.1c it is shown
that a perturbation of a nondispersive solution of the KdV-equation
introduces a dispersion effect, and in Fig.1d, it is shown that this 
results in a different propagation velocity of perturbation.
If the unperturbed solution is a localized non-dispersive function, both
the unperturbed solution and the perturbation travel with a different
velocity depending on the position of the poles of the perturbation. 
The velocity of every spectral component of the perturbation
is determined by its corresponding pole position. In the following we
examine the behavior of the interaction terms.

\subsection{The time-evolution of the interaction terms}
The function $F_{int}(x,k,t)$ corresponding to the interaction term 
contains 
contributions of
both the unperturbed reflection coefficient $R(k,t)$ and
the perturbation of the reflection coefficient $\overline{R}(k,t)$. 
In Fig.6, it
consists of all the diagrams having both solid and open dots. An 
analytic expression of all these diagrams is given by the following 
equation:
\begin{eqnarray}
F_{int}(x,k,t) & = &
\int_{-\infty}^{\infty} \int_{-\infty}^{\infty}
C(k,k^{\prime},t)\overline{C}(k^{\prime},k^{\prime\prime},t) dk^{\prime} 
dk^{\prime \prime}  
+ \int_{-\infty}^{\infty} \int_{-\infty}^{\infty}
\overline{C}(k,k^{\prime},t)C(k^{\prime},k^{\prime\prime},t)dk^{\prime} 
dk^{\prime \prime} 
\nonumber \\
&+& \int_{-\infty}^{\infty} \int_{-\infty}^{\infty}
\int_{-\infty}^{\infty} 
C(k,k^{\prime},t)\overline{C}(k^{\prime},k^{\prime\prime},t)
\overline{C}(k^{\prime\prime},k^{\prime\prime\prime},t)
dk^{\prime} 
dk^{\prime \prime}  
dk^{\prime \prime \prime} + \cdots
\label{f_m}
\end{eqnarray} 
Since, both $R(k,t)$ and $\overline{R}(k,t)$ are rational functions of 
the
wave-number, we can carry out the integrations in Eq.(\ref{f_m})
analytically. If we proceed in a similar manner as for the
non-interaction terms, we obtain the following result for the
time-evolution of the interaction terms:
\begin{eqnarray}
u_{int}(x,t) &=&
8 \sum_{i=1}^{N} \sum_{j=1}^{M} R_{i} \overline{R}_{j}
e^{2i \{ (p_{i}+\overline{p}_{j})x + 4( p_{i}^{3} +
\overline{p}_{j}^{3} ) t \} }
\nonumber \\
&+& 
12 \sum_{i=1}^{N} \sum_{j,l=1}^{M}
\frac{
R_{i} \overline{R}_{j} \overline{R}_{l}
(p_{i} + \overline{p}_{j} + \overline{p}_{l} )
}{
(p_{i}+\overline{p}_{j})(\overline{p}_{j} + \overline{p}_{l})
}
e^{
2i \{ (p_{i} + \overline{p}_{j} + \overline{p}_{l} )x
+ 4 ( p_{i}^{3} + \overline{p}_{j}^{3} + \overline{p}_{l}^{3} )t
\} 
}
\label{ker_ser_m1}
\\ 
&+&
12 \sum_{i,j=1}^{N} \sum_{l=1}^{M}
\frac{
R_{i} R_{j} \overline{R}_{l}
(p_{i} + p_{j} + \overline{p}_{l} )
}{
(p_{i}+p_{j})(p_{j} + \overline{p}_{l})
}
e^{
2i \{ (p_{i} + p_{j} + \overline{p}_{l} )x
+ 4 ( p_{i}^{3} + p_{j}^{3} + \overline{p}_{l}^{3} )t
\} 
}
+ \cdots 
\nonumber
\end{eqnarray}
As witnessed from the previous section, it follows from
Eq.(\ref{ker_ser_m1}) that  the interaction term is large with respect
to the unperturbed solution, if the
unperturbed problem has poles close to the origin in the complex plane.
The analytical structure of Eq.(\ref{part_un}),
Eq.(\ref{part_pert}) and Eq.(\ref{ker_ser_m1}) enables us
to formulate an expression for the amplitude behavior of the
perturbation $f(x,t)$. 
If the perturbation does not have poles close to the origin in the
complex plane and if the perturbation is small with respect to the
unperturbed problem at $t=0$, it follows from Eq.(\ref{part_un})
and Eq.(\ref{part_pert}) that the term $u_{pe}(x,t)$ remains
small with respect to $u_{un}(x,t)$. However, as already concluded  in
the previous section, if the unperturbed solution has singularities, 
the structure of the interaction term
(\ref{ker_ser_m1}) introduces necessarily singularities in the 
time-evolution of the
perturbed problem. 
The physical meaning of $u_{int}(x,t)$  is
visualized in Fig.4. The example given in Fig.4 only differs from the
example given in Fig.3, by the position of the poles and residues that
characterize the unperturbed solution. This implies that the function  
$u_{pe}(x,t)$ in Fig.4 does not differ from that in Fig.3. However, due
to the large amplitudes in Fig.4, $u_{pe}(x,t)$ is small with respect
to $u_{un}(x,t)$. 
In the upper panel of Fig.4, the unperturbed
singular solution is plotted. In the lower panel of Fig.4, the
perturbed solution is plotted. As remarked in Sec.3, in this special
case, the perturbation
consists of two parts. One part propagates over a different
characteristic than $u_{un}(x,t)$. The second part propagates over the
characteristic of $u_{un}(x,t)$  and is 
responsible for large amplitude fluctuations  on the characteristic
of the unperturbed solution. 
It is easy to see that an interaction term as given by  equation 
(\ref{ker_ser_m1}) introduces  large amplitude fluctuations.
If we assume
that the unperturbed solution consist of one pole $(p_{i}=p)$ and 
one residue, 
than we find at the characteristic $x= -4 p^{2} t$:
\begin{equation}
u_{int}(x = -4 p^{2},t) =
8 \sum_{j=1}^{M} R \overline{R}_{j}
e^{
8i \overline{p}_{j} \{ \overline{p}_{j}^{2} - p^{2}  \} t 
}
+ \mbox{h.o.t.}
\end{equation}
From this result it follows that the interaction term $f_{int}(x,t)$ 
introduces fluctuations at the characteristic of the unperturbed
solutions. This result explains the behavior of the perturbed singular
solution
in Fig.4.
It is observed in
this figure that certain spectral components of the noise travel with a
different velocity, and that at the characteristic of the unperturbed
solution large amplitude
fluctuations occur. This is the result of the presence
of the interaction term $u_{int}(x,t)$ which has the same magnitude as
the unperturbed solution $u_{un}(x,t)$.
The interaction
term is large with respect to the unperturbed solution if
$|\overline{R}_{i}| \ll p_{j}$ for all possible $\overline{R}_{i}$ and
$p_{i}$.  If this condition is satisfied, the amplitude of the
perturbation $f(x,t)$ is small  with respect to the amplitude of the
unperturbed solution. 

\section{Numerical example}
In this section the results that are obtained analytically in the
previous sections are illustrated numerically in the case of a
reflection coefficient consisting of an infinite number of poles and
residues. In this section we give an numerical example in to illustrate
the stability of the soliton.
In a discrete representation the 
KdV-equation takes the following form:
\begin{equation}
u_{n}^{i+1} = u_{n}^{i} + \Delta t 
\left\{
\frac{ 
u_{n+3}^{i} - 3 u_{n+1}^{i} + 3 u_{n-1}^{i} - u_{n-3}^{i}
}{
(2 \Delta x)^{3} 
} 
-
\frac{ 6 u_{n}^{i} u_{n+1}^{i} - 6 u_{n}^{i} u_{n}^{i} }{
2 \Delta x
}
\right\}
\label{kdv_dis}
\end{equation} 
In Eq.(\ref{kdv_dis}), the solution of the KdV-equation at  
time $i \Delta t$  and position $n \Delta x $ is given 
by $u_{n}^{i}$. In Eq.(\ref{kdv_dis}), $\Delta t$ represents the
time-step and $\Delta x$ represents the distance between two
grid-points. The discrete KdV-equation (\ref{kdv_dis}) is solved
numerically using the fourth-order Runge-Kutta scheme as given in
ref.\cite{Numres}. 

In the numerical example that follows, the KdV-equation is solved on a
line-segment of a total length of $8 \pi$. The initial condition 
given in Fig.7a consists of the standard soliton which is used in
the previous sections ($\beta=1$ and $d=-1$). The soliton is
contaminated with a noise-function having a maximum amplitude of 
$10 \%$ of
the maximum amplitude of the soliton. In Fig.7b, the soliton after 
a propagation-time of $1.1$ sec is given, it can be seen that
the  contamination has started propagating out of the soliton. 
This process
continues, and at $t=2.2$ sec, the contamination has virtually
propagated out of
the soliton. 

This experiment is a numerical confirmation of the 
experiments of the previous sections. It reflects the case
that small perturbations propagate out of the unperturbed soliton so 
that
the noise separates form the unperturbed solution. This is also the
reason why we observe solitons in nature. In the example of Fig.7
the spectral contents of the perturbation are chosen in such a way
that the soliton and all the spectral components of the perturbation
travel in such a way that the unperturbed solution and the perturbation
separate during the course of time. In a real physical situation
solitons are always contaminated with noise at time $t=0$. If we
observe solitons in nature, it follows from this paper that the
spectral components of the perturbation have a small amplitude for low
frequencies. As a result of this, the soliton and the perturbation
separate during the course of time and the soliton is ``born''. If 
the spectral components of the noise function are  
of the same magnitude than the spectral components of the soliton, the
noise propagates with a similar velocity as the soliton, and hence we
can conclude that no soliton is created.

\section{Conclusions}
In this paper the stability of the KdV-equation is investigated. In
particular, attention is paid to the stability  of localized
solutions. 
Is Sec.2, the effects of nonlinearity and dispersion are
investigated. For non-dispersive solutions of the KdV-equation these 
effects have to be in balance. It is pointed out in Sec.2 
that in the cases of small perturbations the balance the dispersion and
the nonlinearity is disturbed.
As a result of this an additional dispersion
effect is introduced, which generates the ``force'' that separates the
noise from the unperturbed solution.

In Sec.3, inverse scattering techniques are used to formulate an
analytical 
expression which describes the behavior of perturbations of
solutions of the KdV-equation. 
From the examples of Sec.3, it can be concluded  for 
soliton-like solutions of the KdV-equation, that stability implies 
that noise
either propagates out of the unperturbed solution, or that the noise
spreads out so that the amplitude is reduced. 
Furthermore, it is observed for singular solutions that although the
noise is propagating partly out of the unperturbed solution, large
amplitude  variations contaminate the perturbed solution.

This behavior is examined in Sec.4, by expanding the perturbation
into a series-solution. It is concluded in Sec.4 that the behavior
of the perturbation $f(x,t)$ of the KdV-equation is strongly correlated
to the behavior of the unperturbed solution. If the unperturbed problem
has singularities the amplitude of the perturbation is in the same
order of magnitude as the amplitude of the unperturbed problem. This
result explains the large amplitude variations from which the perturbed
singular singular problems suffers. 
Furthermore, in Sec.4, a criterion for the position of
the poles and
residues of the perturbation is given to posses a stable behavior of
the soliton.

Lastly, in Sec.5 we give a numerical example is given to illustrate
the stability of the soliton. The perturbation used in Sec.5 has an
infinite number of poles and residues. We observe that in the
numerical case the same conclusions can be drawn as in cases of
analytical perturbations: the soliton possesses stable behavior if 
the noise is propagating out of the soliton. 

\subsection*{Acknowledgments}
This research was supported by the Netherlands Organization for
Scientific Research (N.W.O.) through the Pionier project PGS 76-144.
This is Geodynamics Research Institute (Utrecht University) publication
96.xxx

\newpage
\section*{Appendix A: The inverse problem for rational reflection 
coefficients}
\renewcommand{\theequation}{\mbox{A-\arabic{equation}}}
\setcounter{equation}{0}
 
In this appendix a brief formulation of the inverse problem for
rational reflection coefficients based upon the formulation of
Sabatier \cite{Sabatier} is given. For a detailed treatment of the 
mathematics
we refer to the book of Chadan and Sabatier \cite{Chadan}.
Our starting-point is the following equation:
\begin{eqnarray}
F_{\pm} (k,x,t) -1 & = &
\frac {1}{2 \pi i} \lim_{ \epsilon \rightarrow 0^{+} }
\int_{- \infty}^{\infty}
\frac{ 1-T (k^{\prime},t)
F_{\mp}(k^{\prime},x,t) }
{ k^{\prime} + k + i \epsilon }  dk^{\prime} 
\nonumber \\
& + &  \frac {1}{2 \pi i} \lim_{ \epsilon \rightarrow 0^{+} }
\int_{- \infty}^{\infty}
\frac{     R_{\pm} (k^{\prime},t)
F_{\pm}(k^{\prime},x,t) exp[ \pm 2 ik^{\prime} x ]    }
{ k^{\prime} + k + i \epsilon }  dk^{\prime} 
\label{eq:invcau.a_a}
\end{eqnarray}
In Eq.(\ref{eq:invcau.a_a}) $F_{+}(k,x)$ is defined for 
$ x > 0 $, 
and
$F_{-}(k,x,t)$ for $ x < 0 $, $k \in \cn$.
The function $F_{\pm}(k,x,t)$ is defined by:
\begin{equation}
F_{\pm}(k,x,t)=exp[ {\mp} ikx] f_{\pm}(k,x,t)
\label{eq:fad.a_a}
\end{equation}
The Jost solutions $f_{\pm}(k,x,t)$ are those solutions of the
Schr\"{o}dinger equation satisfying the following boundary conditions:
\begin{eqnarray}
& f_{+}(k,x,t): &   \lim_{x \rightarrow \infty} e^{-ikx}
f_{+}(k,x,t)=1 
\label{eq:jostright.a_a}  \\
& f_{-}(k,x,t): &   \lim_{x \rightarrow - \infty} e^{ikx}
f_{-}(k,x,t)=1
\label{eq:jostleft.a_a}
\end{eqnarray}
They satisfy the following integral equations:
\begin{eqnarray}
f_{+}(k,x,t) & =  & e^{ikx} - \int_{x}^{\infty}
\frac { \sin k(x-y) }{k} V(y,t) f_{+}(k,y,t)dy 
\label{eq:josrrint.a_a}  \\
f_{-}(k,x,t) & = & e^{-ikx} - \int_{-\infty}^{x}
\frac { \sin k(x-y) }{k} V(y,t) f_{-}(k,y,t)dy
\label{eq:jostlint.a_a}
\end{eqnarray}
It is well known that the functions $f_{\pm}(k,x,t)$
and therefore also the functions $F_{\pm}(k,x,t)$ are holomorphic in 
$\cn^{+}$ \cite{Chadan}.
The potential $V(x,t)$ has to be in the Faddeev class $L_{1}^{1}$:
\begin{equation}
\int_{ - \infty}^{\infty} (1+|x|)|V(x,t)| < \infty
\label{eq:fadclass.a_a}
\end{equation}
The scattering coefficients $R_{+}(k,t), R_{-}(k,t), T(k,t)$ are defined by 
the asymptotic
behavior of the physical solution of the Schr\"{o}dinger equation:
\begin{equation}
\psi_{1}(k,x,t) \sim 
\left\{
\begin{array}{ll}
e^{ikx}+R_{+}(k,t)e^{-ikx} & x < 0 \\
T(k,t)e^{ikx}              & x \rightarrow + \infty
\end{array}
\right.
\label{eq:asympl.a_a}
\end{equation}
\begin{equation}
\psi_{2}(k,x,t) \sim 
\left\{
\begin{array}{ll}
T(k,t)e^{-ikx}             & x < 0   \\
e^{-ikx}+R_{-}(k,t)e^{ikx} & x \rightarrow + \infty 
\end{array}
\right.
\label{eq:asympr.a_a}
\end{equation}
In the case of rational reflection coefficients they take the following
form \cite{Sabatier}:
\begin{eqnarray}
R_{+}(k,t=0) & = & \frac{  P(-k)  }{  \prod_{j=1}^{q} (\lambda_{j} - k )  }
\prod_{ \mu_{i} \in M^{+} } \frac{ \mu_{i} + k }{ \mu_{i} - k } 
\prod_{ \lambda_{l} \in L^{+} } \frac{ \lambda_{i} + k }
{ \lambda_{i} - k }  
\label{eq:regenl.a_a} \\
T(k,t=0) & = &  \frac 
{ \prod_{i=1}^{q}(\mu_{i} + k ) }
{ \prod_{j=1}^{q}(\lambda_{i} + k ) } 
\label{eq:transgen.a_a}  \\
R_{-}(k,t=0) & = &  \frac{ P(k) }{ \prod_{j=1}^{q} (\lambda_{j} - k )}
\prod_{ \mu_{i} \in M^{-} } \frac{ \mu_{i} - k }{ \mu_{i} + k } 
\prod_{ \lambda_{l} \in L^{-} } \frac{ \lambda_{i} - k }
{ \lambda_{i} + k }   
\label{eq:regenr.a_a}
\end{eqnarray}
Following Sabatier \cite{Sabatier}, the degree of the polynomial 
$P(k)$ has to be smaller than $q$. Further, Im $ \mu_{i} > 0 $ except if
$\mu_{i} = 0 $, Im $ \lambda_{l} < 0$. The transmission coefficient
$T(k,t)$ is supposed to be an irreducible fraction, and the sets $M^{+}$,
$M^{-}$, $L^{+}$, $L^{-}$ contain numbers $ \neq 0$.
If the potential is real then  $\mu_{k}$,
$\lambda_{k}$ are pure imaginary. If $\mu_{k}$,$\lambda_{k}$ are not
pure imaginary then there exists 
$ - \mu_{k}^{\ast}$, 
$ - \lambda_{k}^{\ast}$.
It can be shown that $T(k)$ is meromorphic in $\cn^{+}$
and if there are poles they are in Im $k$ \cite{Chadan}.
If there are no bound
states, $T(k,t)F_{\mp}(k,x,t)$ is holomorphic in $\cn^{+}$ and the first
integral of (\ref{eq:invcau.a_a}) is zero. If $T(k,t)F_{\mp}(k,x,t)$ is
holomorphic in
$\cn^{+}$ and all the poles $p_{i}$ of $R_{+}(k,t)$ are simple,
the
integral (\ref{eq:invcau.a_a}) can be solved by contour integration in 
the upper-half 
plane.  The result is:
\begin{equation}
F_{+} (k,x,t) -1  = 
\sum_{p_{j} \in {\cal P } }
\frac{ R_{j} 
F_{+}(p_{j},x,t) e^{2i(p_{j}x -4 p_{j}^{3} t)}    }
{ p_{j} + k }  
\label{eq:invrat.a_a}
\end{equation}
The time-evolution of the residues that is used in equation
(\ref{eq:invrat.a_a}), is given in ref.\cite{Dorren0}.
Equation (\ref{eq:invrat.a_a}) can be solved by letting take $k$ the
values of the discrete poles $p \in {\cal P}$. We then obtain  
a linear set of
algebraic equations that determine $F_{+}(p_{j},x,t)$ for all values of
$j$. This set of equations can be solved making use of Cramer's rule. 
We obtain after resubstituting the result in equation
(\ref{eq:invrat.a_a}), using equation (\ref{kertran_c}) and putting 
$x=y$:
\begin{equation}
K_{+}(x,x,t)=
\frac{ {\cal D}_{+}^{\prime}(x,t) }{  {\cal D}_{+}(x,t)   }
\label{eq:kernel.a_a}
\end{equation}
where:
\begin{equation}
{\cal D}_{+}(x,t) = det \{ \delta_{ij} - ( p_{i} + p_{j} )^{-1}
R_{j} e^{ 2i(p_{j}x-4p_{j}^{3}t) } \}
\label{eq:det.a_a}
\end{equation}  
and ${\cal D}_{+}^{\prime}(x,t)$ is the derivative of ${\cal D}_{+}(x,t)$ with
respect to $x$.

\newpage
\section*{Captions for figures}

\noindent
{\bf Figure 1}: A: Unperturbed soliton (solid line) and perturbed
soliton (dashed line). B: the dispersion of the unperturbed soliton
(short-dashed), the nonlinearity of the unperturbed soliton
(long-dashed line) and the sum of the dispersion and nonlinearity
(solid line). C: Similar as for Fig.1b, but now for the perturbed
soliton. D: The time-derivative of the perturbation. E:
Finite-difference solution of the perturbed soliton at $t=2.2$
sec.

\vspace{.5cm}
\noindent
{\bf Figure 2}:
A: Singular solution (solid line), the dispersion (short-dashed
line) and the nonlinearity (long-dashed line). B: The dispersion is
case of a  $10 \%$ amplitude error (short-dashed line). The
nonlinearity in case of a $10 \%$ amplitude error (long-dashed line)
and the sum of the nonlinearity and the dispersion (solid-line).

\vspace{.5cm}
\noindent
{\bf Figure 3}: A: Time-evolution  of a soliton in case of a 
perturbation consisting of nine additional poles and 
residues (lower panel) and the time-evolution of the unperturbed
soliton, (upper panel). B: The initial condition (short-dashed),a
the solution at 
$t=1$ (long-dashed line) and the
unperturbed soliton (solid-line).

\vspace{.5cm}
\noindent
{\bf Figure 4}: Time-evolution is case of a singular solution
contaminated  with the same perturbation as in the previous figure
(lower panel) and the time-evolution of the unperturbed singular
solution (upper panel). 

\vspace{.5cm}
\noindent
{\bf Figure 5}: Dyson's representation for the integral equation
(\ref{fexp_c})

\vspace{.5cm}
\noindent
{\bf Figure 6}: Dyson's representation of the series expansion of
Fig. 5.

\vspace{.5cm}
\noindent
{\bf Figure 7}: A: Plot of the soliton contaminated with a perturbation 
consisting of an infinite number of poles and residues. B: The initial
condition, the solution at t=1.1 sec and the solution at time $t=2.2$
sec.

\end{document}